\pdfoutput=1

\documentclass[11pt]{article}

\usepackage[final]{acl}

\usepackage{times}
\usepackage{latexsym}

\usepackage[T1]{fontenc}

\usepackage[utf8]{inputenc}

\usepackage{microtype}

\usepackage{inconsolata}

\usepackage{graphicx}
\usepackage{subcaption}
\usepackage{amsmath}
\usepackage{mathtools}
\usepackage{comment}
\usepackage{makecell}
\usepackage{placeins}
\usepackage{balance}
\title{Breaking Down Power Barriers in On-Device Streaming ASR: \\Insights and Solutions}

\author{
    \textbf{Yang Li}\thanks{Co-first authors.}\thanks{Corresponding author (\texttt{jerryyangli@gmail.com}). Work partially done while employed at Meta and partially while at Iowa State University.}$^{\,2}$
    \quad\,
    \textbf{Yuan Shangguan}\footnotemark[1]\thanks{Work done while employed at Meta.}$^{\,3}$
    \quad
    \textbf{Yuhao Wang}$^{1}$
    \quad\,
    \textbf{Liangzhen Lai}$^{1}$
    \\
    \textbf{Ernie Chang}$^{1}$
    \quad
    \textbf{Changsheng Zhao}$^{1}$
    \quad
    \textbf{Yangyang Shi}$^{1}$
    \quad 
    \textbf{Vikas Chandra}$^{1}$
    \\
    \\
    $^{1}$Meta
    \quad
    $^{2}$Iowa State University
    \quad
    $^{3}$Google
}

\begin{document}
\maketitle
\begin{abstract}
Power consumption plays a crucial role in on-device streaming speech recognition, significantly influencing the user experience. This study explores how the configuration of weight parameters in speech recognition models affects their overall energy efficiency. We found that the influence of these parameters on power consumption varies depending on factors such as invocation frequency and memory allocation. Leveraging these insights, we propose design principles that enhance on-device speech recognition models by reducing power consumption with minimal impact on accuracy. Our approach, which adjusts model components based on their specific energy sensitivities, achieves up to 47\% lower energy usage while preserving comparable model accuracy and improving real-time performance compared to leading methods.
\end{abstract}

\section{Introduction\label{sec:intro}}
Streaming \underline{a}utomatic \underline{s}peech \underline{r}ecognition (streaming ASR) enables real-time transcription of speech to text with latency typically under 500 milliseconds, supporting applications such as interface navigation, voice commands, real-time communication, and accessibility on mobile and wearable devices. However, high power consumption poses a significant challenge, limiting usability by requiring frequent recharges. Improving the energy efficiency of on-device streaming ASR is therefore essential for enhancing user experience.

We focus on on-device streaming ASR models, particularly the Neural Transducer \cite{graves2012sequence}, which combines an Encoder for acoustic modeling, a Predictor for language modeling, and a Joiner to integrate their outputs (see Figure~\ref{fig:rnnt}). Widely regarded as the standard for on-device streaming ASR~\cite{graves2013speech, streamingE2E, li2021better}, the Neural Transducer excels in balancing computational efficiency and accuracy. We train and evaluate over 180 Neural Transducer models\footnote{Traning each model requires 640-960 V100 GPU hours.}, exploring architectures including Emformer~\cite{Emformer} and Conformer~\cite{conformer} while varying component sizes. This extensive study reveals how the components impact accuracy, real-time factor (RTF),\footnote{ RTF is the ratio of inference time to the speech segment duration, with lower values indicating faster processing.} and power consumption.

Our analysis reveals several key findings: (1)~Energy usage in streaming ASR models is driven by memory traffic for loading weights, which depends on the invocation frequency of components and their memory hierarchy placement. (2)~Invocation frequencies vary widely, with the Joiner being called far more often than the Predictor, and the Predictor more than the Encoder. Despite comprising only 5–9\% of the model's size, the Joiner accounts for 48–73\% of its power consumption. (3)~We identify an exponential relationship between model accuracy and encoder size, suggesting new directions for streaming ASR research.

Building on these insights, we propose a targeted compression strategy to optimize energy efficiency with minimal accuracy loss. This approach evaluates power and accuracy sensitivity for each component, prioritizing compression of components with higher power sensitivity and lower accuracy sensitivity. Specifically, we focus on compressing the Joiner first, followed by the Predictor and Encoder, and aim to store the Joiner’s weights in energy-efficient local memory. Experiments on LibriSpeech~\cite{librispeech} and Public Video datasets show our method reduces energy usage by up to 47\% and lowers RTF by up to 29\%, while maintaining comparable accuracy to state-of-the-art compression strategies. Unlike previous approaches, our method effectively leverages the diverse runtime characteristics of ASR components, showcasing its superior efficiency.

This paper makes the following contributions:
\begin{itemize}
\item Power consumption analysis: We reveal that ASR component energy usage depends not only on model size but also on invocation frequency and memory placement. This challenges the prevailing belief that larger components inherently consume more energy, emphasizing the role of operational dynamics and memory management.
\item Energy-efficient design: We propose design guidelines that reduce energy consumption by up to 47\% and RTF by up to 29\% while maintaining comparable model accuracy to state-of-the-art methods.
\item Accuracy-size relationship: We uncover an exponential relationship between model accuracy and encoder size, showing diminishing gains with larger encoders and advocating for more efficient use of computational and memory resources in on-device streaming ASR.
\end{itemize}

An earlier version of this paper was released as a preprint on arXiv \cite{li2024weightscreatedequalenhancing}.

\section{Background\label{sec:background}}
\subsection{On-Device Streaming ASR\label{subsec:transducer}}

The Neural Transducer, introduced in~\cite{graves2012sequence}, is the state-of-the-art solution for on-device streaming speech recognition~\cite{graves2013speech, streamingE2E, li2021better}. It aligns audio and text~\cite{E2ESurvey} by integrating a compact language model and acoustic model within a single framework, making it ideal for resource-constrained devices due to its reduced memory footprint~\cite{shangguan2019optimizing, memoryEfficienRNNT}. With sub-500 millisecond latency, it meets the demands of streaming applications, and it is widely adopted by leading companies for on-device ASR~\cite{li2023folding, le2023factorized, wang2023multi, radfar2022convrnn}.

\begin{figure}
\centering
\includegraphics[width=1.0\linewidth]{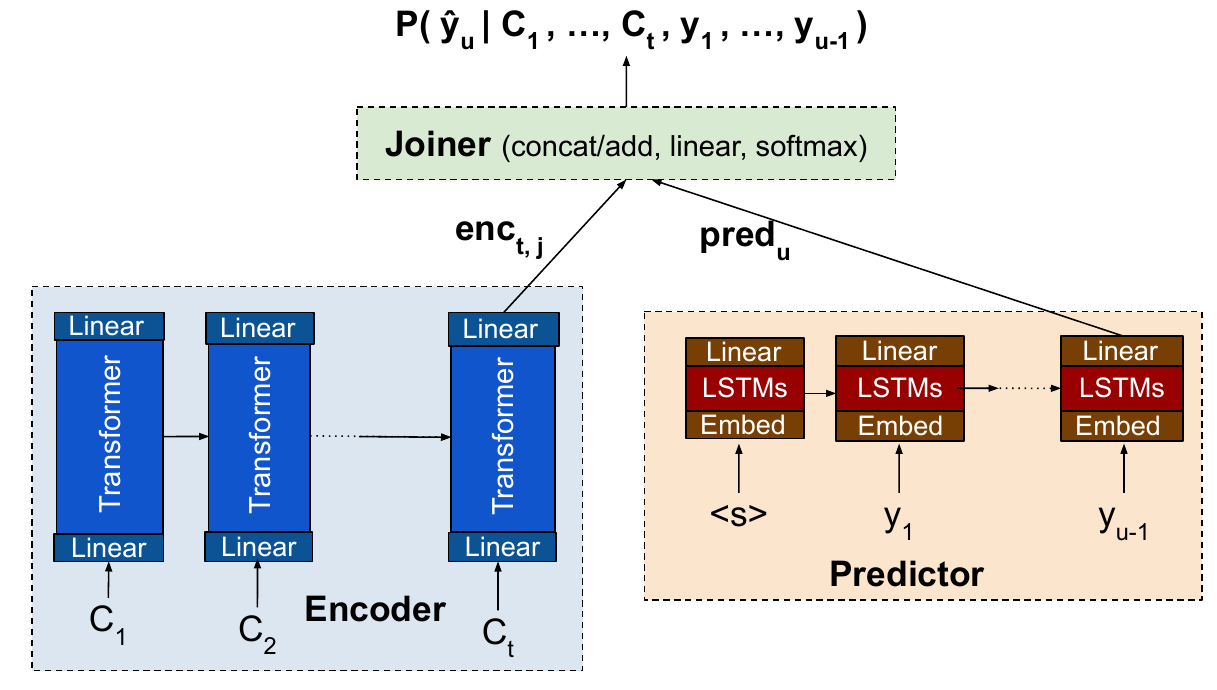}
\caption{A schematic representation for the Transformer-based Neural Transducer.}
\label{fig:rnnt}
\end{figure}

The architecture comprises three components: an Encoder, a Predictor, and a Joiner (Figure~\ref{fig:rnnt}). The Encoder processes chunks of audio ($C_1$, ..., $C_t$), each consisting of frames ($\mathbf{x}_{t, 1}$, ..., $\mathbf{x}_{t, n}$) with 80-dimensional log Mel-filterbank features derived from a 25 ms sliding window with a 10 ms step. The Encoder maps frames to embeddings ($\mathbf{enc}_{t, j}$). The Predictor uses previously predicted tokens (${y}_{1}$, ..., ${y}_{u-1}$) to forecast the embedding of the next token ($\mathbf{pred}_{u}$). The Joiner combines the embeddings from the Encoder and Predictor, processes them through a feedforward network, and applies a softmax to generate the probability distribution over sentence-piece targets and a "blank" token indicating the end of a frame's transcription.

Recent studies~\cite{Emformer, moritz2020streaming, DongICASSP18, ZhangICASSP20, YehArxiv2019, conformer, WangICASSP20, KaritaASRU19} show a preference for Transformer-based Encoders in Neural Transducers. We implement the Encoder using Emformer~\cite{Emformer} and Conformer~\cite{conformer}, two Transformer variants optimized for streaming. These designs enable chunk-based frame processing, reducing Encoder invocation frequency compared to the Predictor and Joiner, which process frames individually. The Predictor is invoked per meaningful output token, while the Joiner operates for both meaningful tokens and frequent "blank" tokens. This results in a hierarchy of invocation frequency: the Joiner is used most, followed by the Predictor, and then the Encoder.




\begin{figure}[b]
\centering
\includegraphics[width=0.65\linewidth]{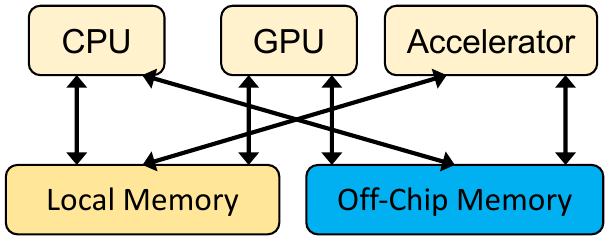}
\caption{Architecture of mobile and wearable devices.}
\label{fig:device_arch}
\end{figure}

\subsection{Mobile and Wearable Devices\label{subsec:computing_device}}

\begin{figure*}[t]
    \centering
    \begin{subfigure}{.27\textwidth}
        \raggedright
        \includegraphics[width=\linewidth]{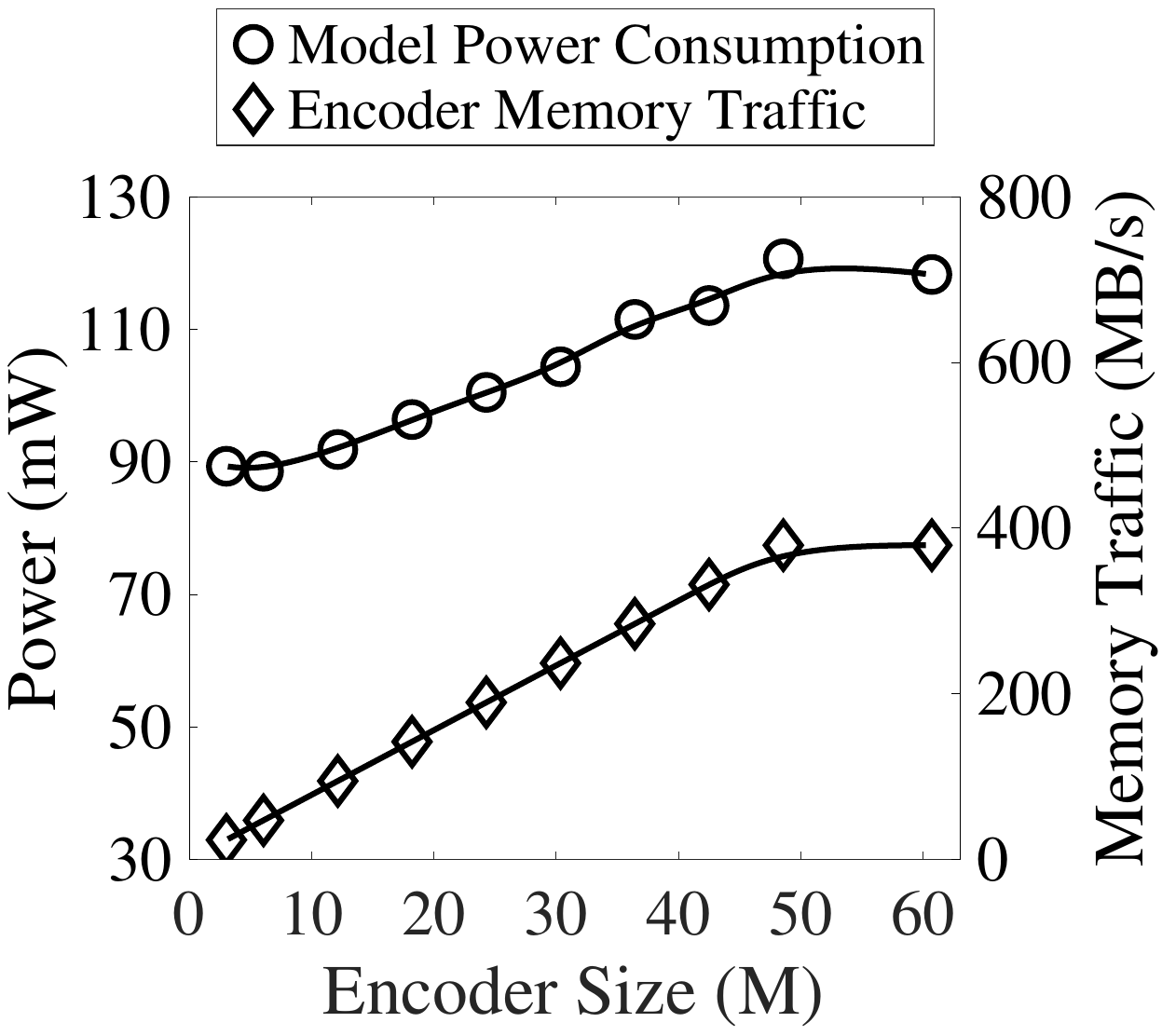}
        \caption{Compressing Encoder}
    \end{subfigure}\hfill%
    \begin{subfigure}{.27\textwidth}
        \centering
        \includegraphics[width=\linewidth]{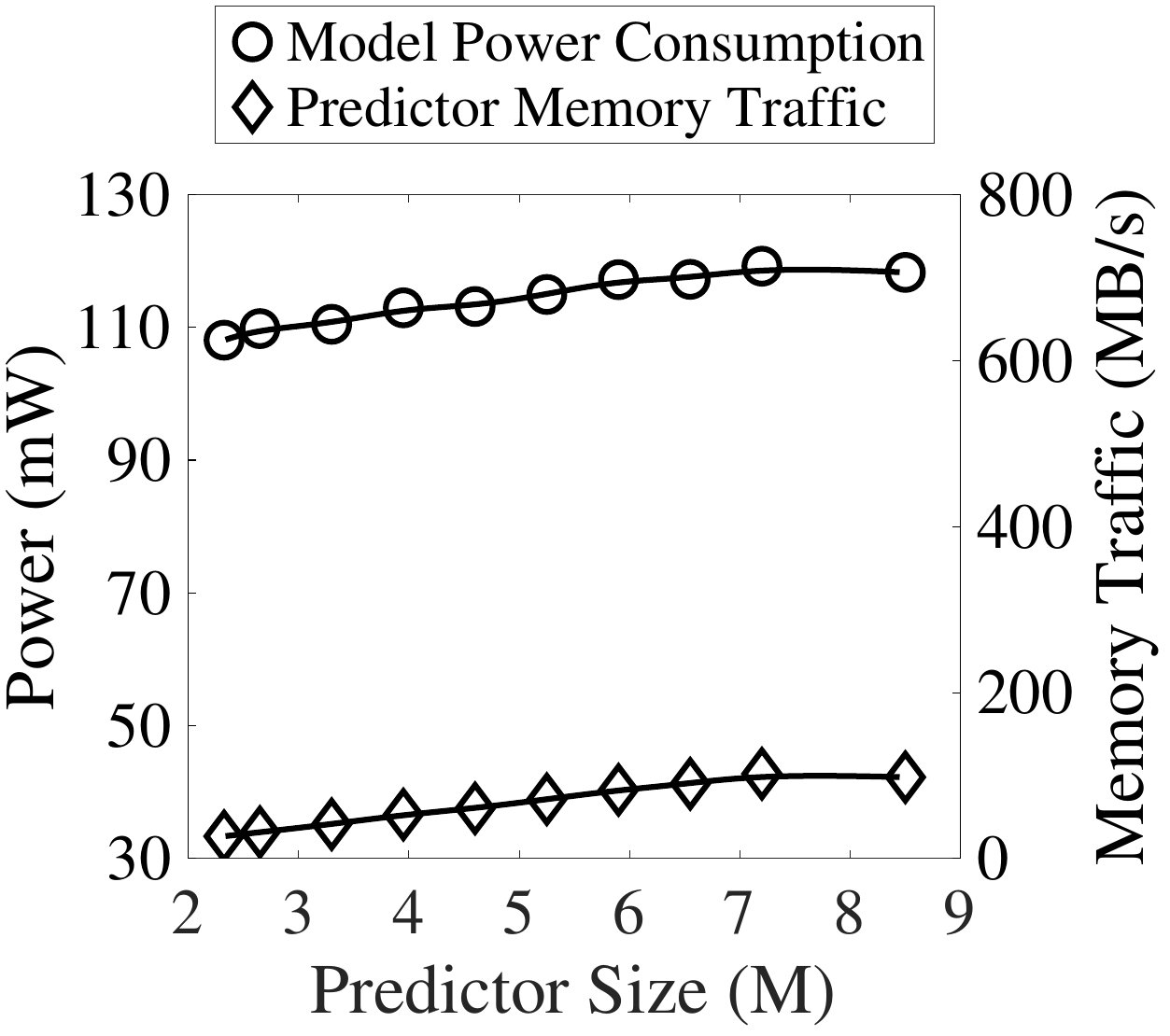}
        \caption{Compressing Predictor}
    \end{subfigure}\hfill
    \begin{subfigure}{.27\textwidth}
        \raggedleft
        \includegraphics[width=\linewidth]{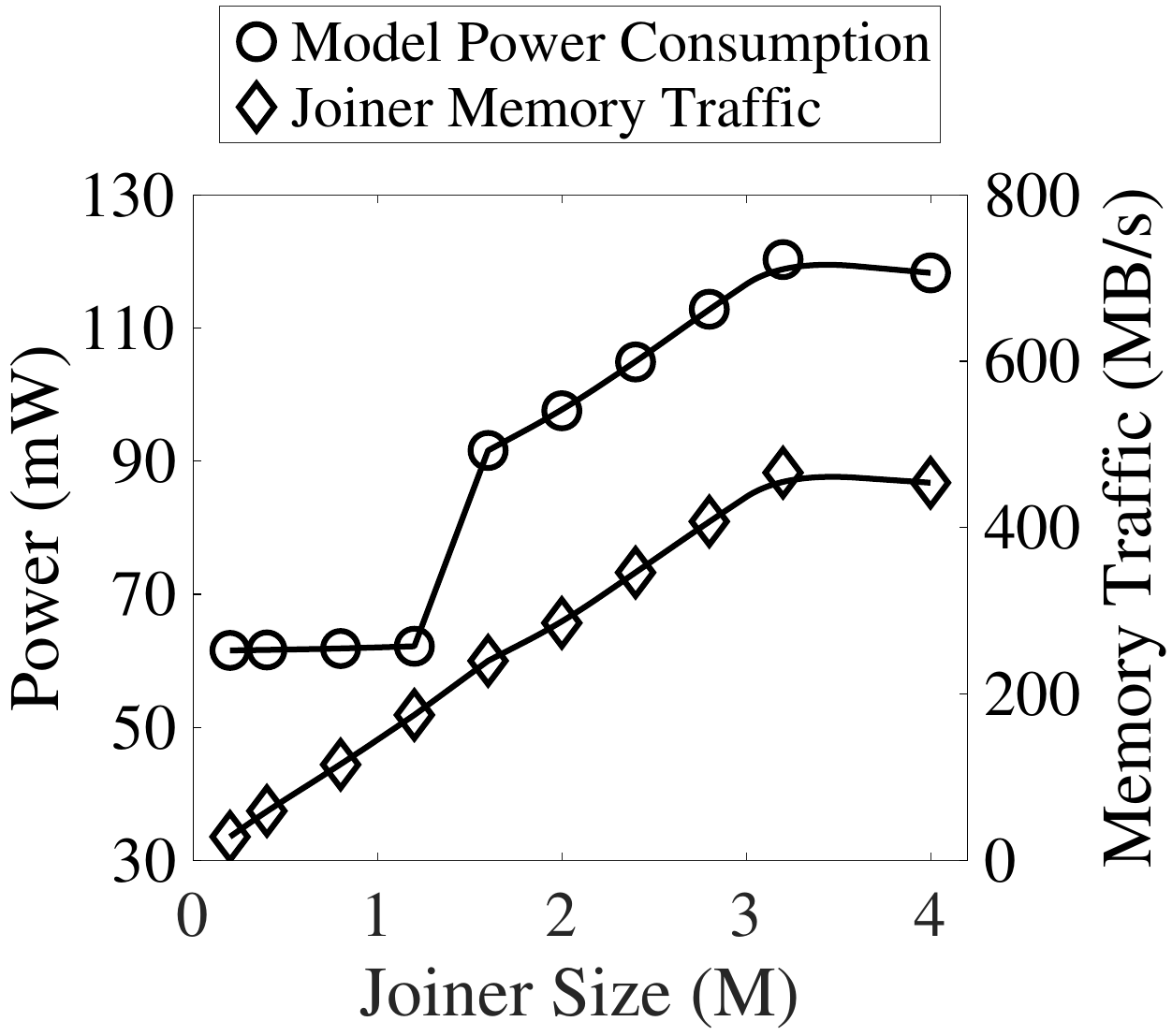}
        \caption{Compressing Joiner}
    \end{subfigure}
    \caption{Models trained on LibriSpeech: Model power consumption with compressing an individual component (Encoder, Predictor, or Joiner) while keeping the sizes of the other two components constant.}
    \label{fig:power_analysis}
\end{figure*}

As shown in Figure~\ref{fig:device_arch}, mobile and wearable devices feature processors such as mobile CPUs, GPUs, and hardware accelerators, all optimized for energy efficiency. For example, a neural network accelerator highlighted by \cite{UNPU} achieves 5 GOPS/mW (INT8), consuming just 1 mW for 5 billion INT8 operations per second. These processors interact with two memory types: \textit{local memory} (e.g., SRAM, eDRAM, on-chip DRAM) and \textit{off-chip memory} (e.g., DRAM). Local memory offers faster, energy-efficient access, with 64-byte read/write operations taking 0.5–20 ns and consuming 1.1–1.5 pJ/byte~\cite{On_Chip_Mem}. In contrast, off-chip memory is slower and less efficient, with 64-byte operations taking 50–70 ns and using about 120 pJ/byte~\cite{On_Chip_Mem}.
This stark energy efficiency gap makes memory operations a dominant energy drain in on-device streaming ASR.

\begin{table}
\vspace{0.3em}
\centering
\renewcommand{\arraystretch}{1.0}
\resizebox{1.0\columnwidth}{!}{
\begin{tabular}{l|c c c}
\hline
 & Encoder & Predictor & Joiner \\ \hline
Size (M) & 60.70 & 8.50 & 4.00 \\ \hline
Compute Power (mW) & 0.80 & 0.03 & 0.19 \\ \hline
Memory Power (mW) & 47.78 & 12.33 & 57.13 \\ \hline
Invocation Frequency (Hz) & 6.25 & 11.53 & 113.50 \\ \hline
\end{tabular}}
\caption{A typical model trained on LibriSpeech.}
\label{table: libri_model}
\end{table}

In our study, we ran streaming ASR models on a Google Pixel-5 smartphone, measuring RTF and workload statistics including the number of operations and component invocations. These workload metrics remain consistent across device platforms. Therefore, the power analysis derived from these metrics applies broadly to other mobile and wearable devices. 
We modeled ASR power consumption using established methodologies~\cite{li2023folding, micron06, UHMEM, isca09}, leveraging computing and memory power parameters from authoritative literature in the circuits community~\cite{UNPU, On_Chip_Mem}. Our setup includes a hardware accelerator, 2 MB of local memory (1.5 MB for weights and 0.5 MB for activations), and 8~GB of off-chip memory, with local memory treated as a scratchpad for flexible allocation. This setup does not represent a specific commercial hardware platform or product; rather, it serves as a general model that is broadly representative of most mobile and wearable devices.

\section{Power and Accuracy Analysis of On-Device Streaming ASR\label{sec:analysis}}

In this section, we use Adam-pruning \cite{yang2022omni}, a state-of-the-art weight pruning technique for speech recognition,\footnote{Adam-pruning is detailed in Appendix \ref{sec:appendix_adamprune}.} to adjust the sizes of the Encoder, Predictor, and Joiner in ASR models. This generates ASR models of varying sizes, enabling analysis of their power consumption and accuracy, yielding key insights.


\subsection{Power Analysis}
Table~\ref{table: libri_model} summarizes the characteristics of a typical on-device streaming ASR model trained on LibriSpeech~\cite{librispeech}, including size, component invocation frequency, computing power, and memory power. The data reveals that computing power accounts for less than 1\% of total power, with memory power dominating due to frequent weight loading. Although the Encoder holds over 83\% of the weights, the Joiner, invoked 18 times more often, generates 1.2 times more memory traffic and consumes more power. This challenges the prevailing belief that larger components consume more energy, highlighting the importance of operational dynamics in energy optimization.

Figure~\ref{fig:power_analysis} examines power consumption by compressing individual components (Encoder, Predictor, or Joiner) while keeping the others unchanged. The results show that power closely tracks memory traffic, which depends on component size and invocation frequency. Notably, compressing the Joiner below 1.2M parameters does not reduce power further, as its weights then fit into energy-efficient local memory, minimizing data-loading energy costs. This underscores the strategic advantage of placing the most energy-intensive components in local memory to optimize energy efficiency.

We also investigate the effects of input stride and chunk size—two key hyperparameters of streaming ASR—on the model's power consumption, revealing some interesting observations. Detailed results are provided in Appendix \ref{sec:appendix_param_sweep}.

\begin{figure*}
    \centering
    \begin{subfigure}{.24\textwidth}
        \raggedright
        \includegraphics[width=\linewidth]{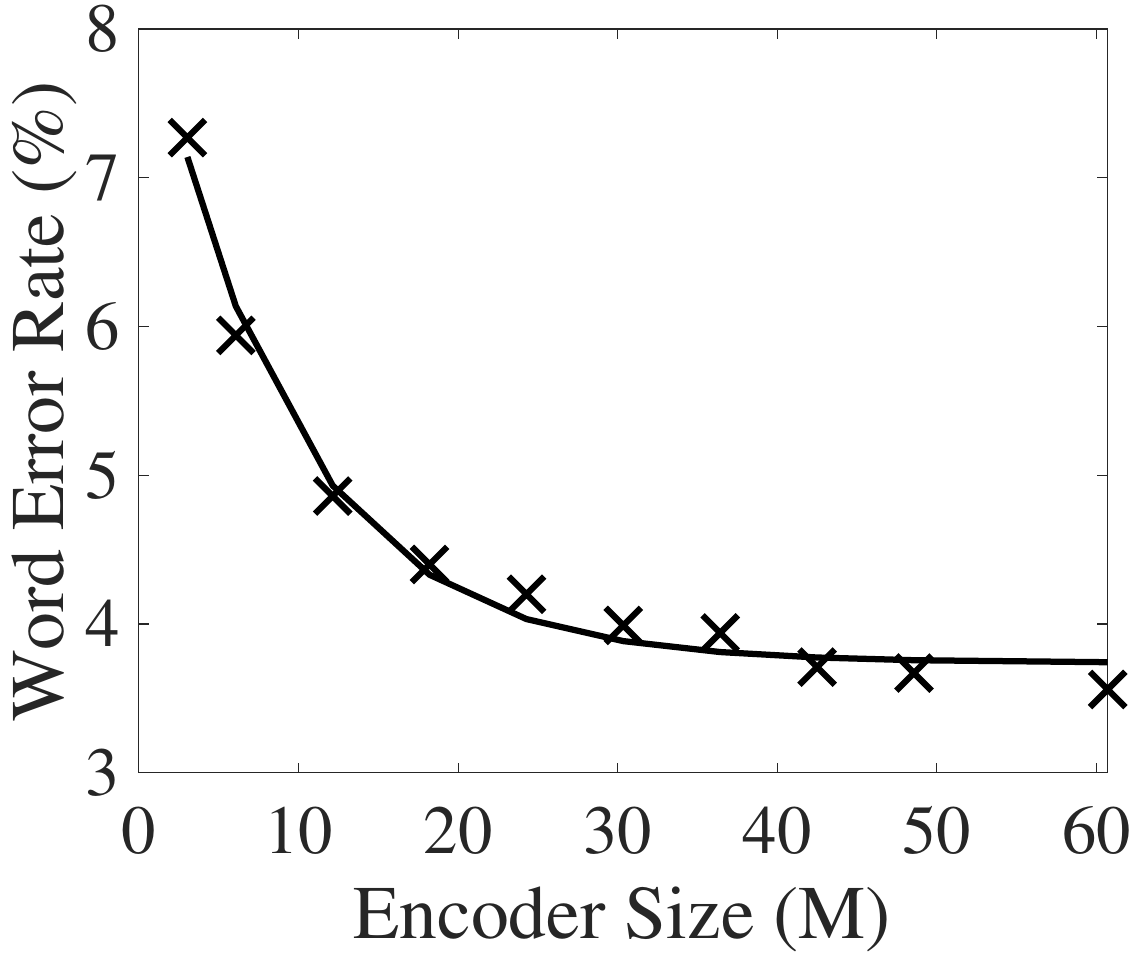}
        \caption{Compressing Encoder}
    \end{subfigure}\hfill%
    \begin{subfigure}{.24\textwidth}
        \centering
        \includegraphics[width=\linewidth]{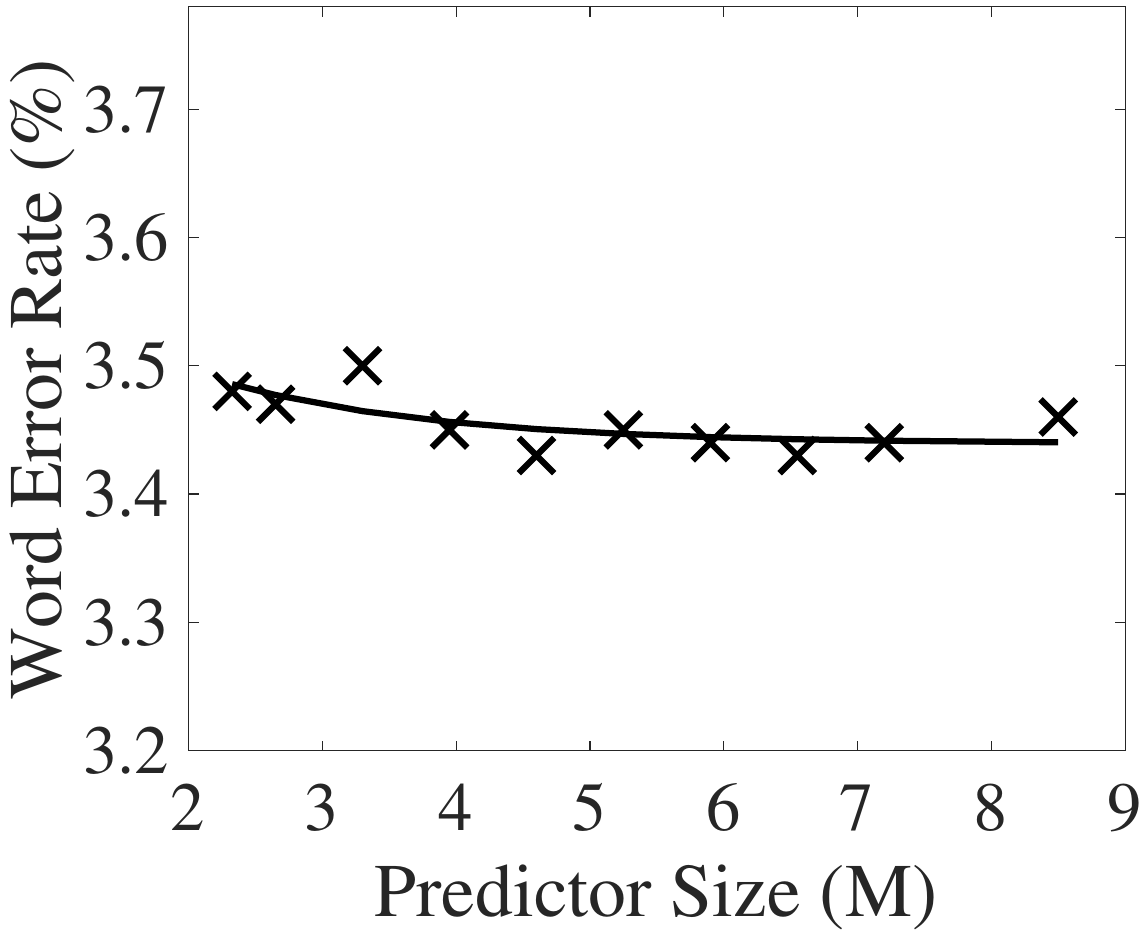}
        \caption{Compressing Predictor}
    \end{subfigure}\hfill
    \begin{subfigure}{.24\textwidth}
        \raggedleft
        \includegraphics[width=\linewidth]{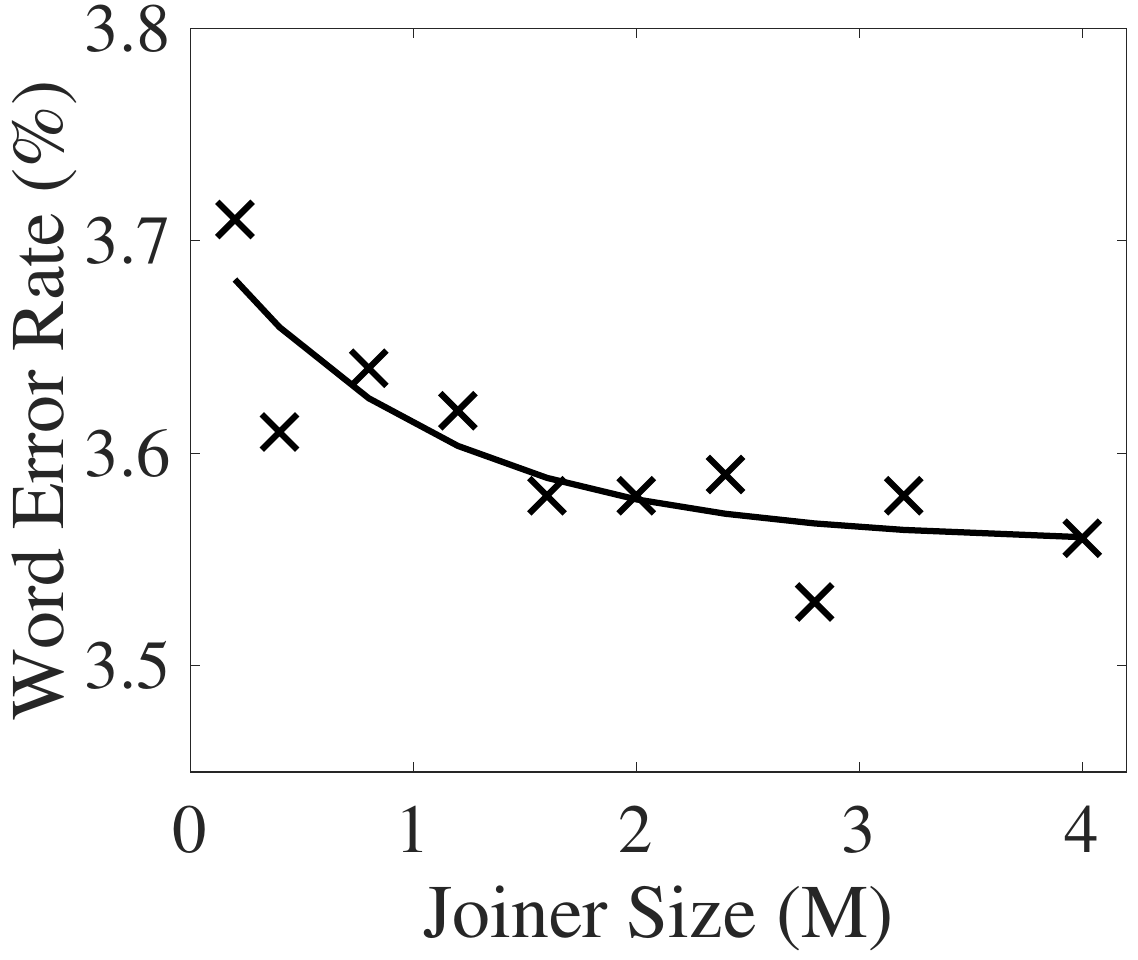}
        \caption{Compressing Joiner}
    \end{subfigure}
    \caption{Models trained on LibriSpeech: Word error rate on Test-Clean with compressing an individual component (Encoder, Predictor, or Joiner) while keeping the sizes of the other two components constant.}
    \label{fig:test_clean_analysis}
\end{figure*}
\begin{figure*}[t]
    \centering
    \begin{subfigure}{.24\textwidth}
        \raggedright
        \includegraphics[width=\linewidth]{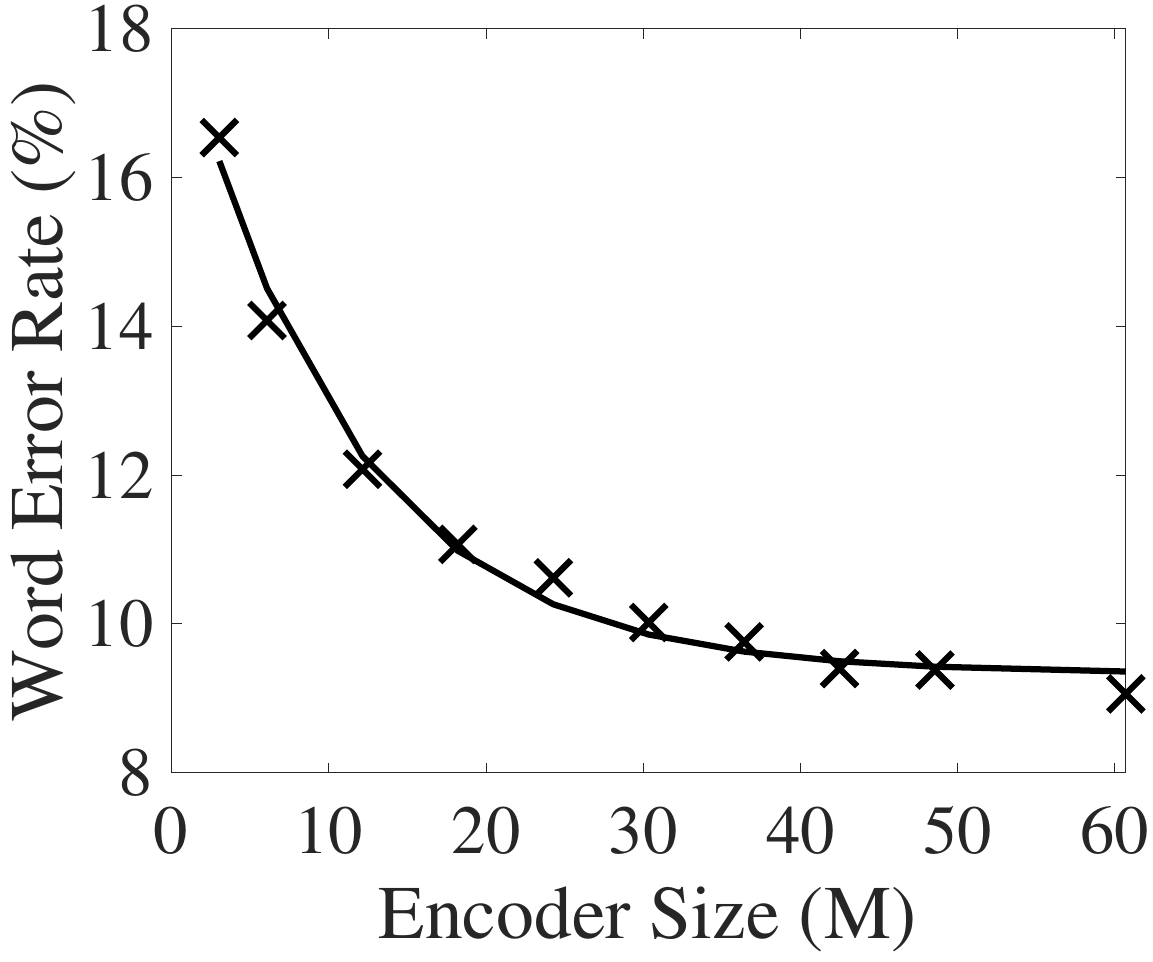}
        \caption{Compressing Encoder}
    \end{subfigure}\hfill%
    \begin{subfigure}{.24\textwidth}
        \centering
    \includegraphics[width=\linewidth]{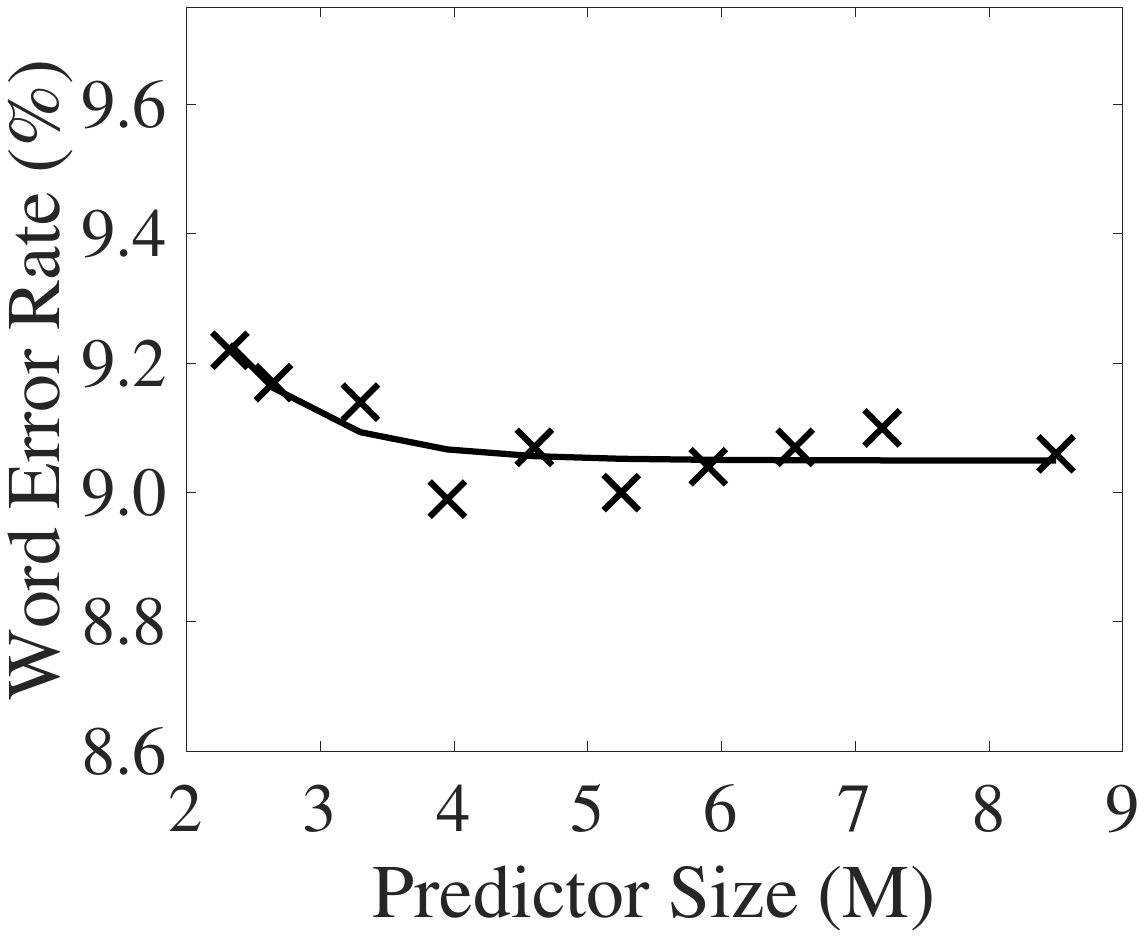}
        \caption{Compressing Predictor}
    \end{subfigure}\hfill
    \begin{subfigure}{.24\textwidth}
        \raggedleft
    \includegraphics[width=\linewidth]{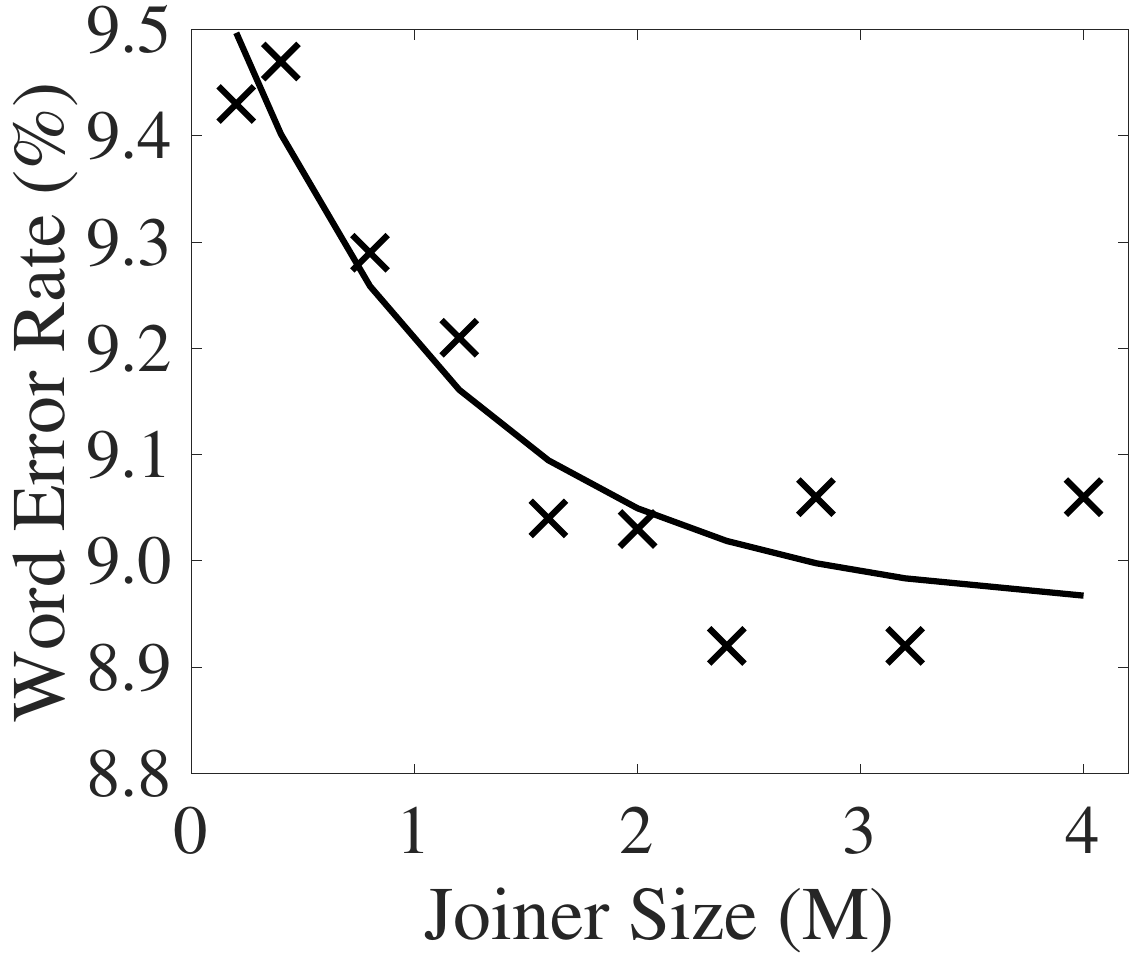}
        \caption{Compressing Joiner}
    \end{subfigure}
    \caption{Models trained on LibriSpeech: Word error rate on Test-Other with compressing an individual component (Encoder, Predictor, or Joiner) while keeping the sizes of the other two components constant. }
    \label{fig:test_others_analysis}
\end{figure*}

\vspace{0.1em}\subsection{Accuracy Analysis}
Figures~\ref{fig:test_clean_analysis} and \ref{fig:test_others_analysis} show the word error rates for compressed models on LibriSpeech's test-clean and test-other sets. Reducing component sizes generally increases word error rates.\footnote{Variability in Predictor and Joiner compression curves stems from randomness in training and pruning.} Among the components, the Predictor is least sensitive to compression, indicating that using a smaller Predictor or omitting it entirely has minimal impact on accuracy. In contrast, the Encoder and Joiner are more sensitive to compression, with encoder size showing an exponential relationship to word error rate:
\begin{equation}
\scalebox{0.85}{$
\text{Word Error Rate} = \exp\left( a \cdot \text{encoder\_size} + b \right) + c
\label{eq:1}
$}
\end{equation}
Fitting this function yielded parameters $a$, $b$, and 
$c$ with adjusted R-squared values of 0.9832 (test-clean) and 0.9854 (test-other), confirming the model's strong fit. Similar trends were observed in other datasets (Appendix~\ref{sec:appendix_product_asr}). This exponential relationship suggests diminishing returns with larger encoder sizes, encouraging the community to rethink encoder design in ASR systems.

\vspace{0.5em}\section{ASR Energy Efficiency Optimization \label{sec:guideline}}
We aim to minimize the power consumption of streaming ASR models with minimal performance impact by evaluating the \textbf{power} and \textbf{accuracy sensitivities} of the Encoder, Predictor, and Joiner components. These sensitivities quantify the change in power consumption and performance, respectively, for a unit reduction in component size:
\begin{equation}
\scalebox{0.85}{$
\begin{aligned}
&\text{Power Sensitivity}_{\text{component}} \coloneqq \frac{\Delta \text{Power}}{\Delta \text{Size}_{\text{component}}} \\[1em]
&\text{Accuracy Sensitivity}_{\text{component}} \coloneqq \frac{\Delta \text{Accuracy}}{\Delta \text{Size}_{\text{component}}}
\end{aligned}
$}
\end{equation}
Here, \text{component} refers to the Encoder, Predictor, or Joiner, and accuracy is inversely related to the word error rate.

The power consumption of on-device streaming ASR is primarily due to loading model weights from memory. Power sensitivity is therefore expressed as:
\begin{equation}
\scalebox{0.85}{$
\begin{aligned}
&\text{Power Sensitivity}_{\text{component}} \\
&= \frac{\Delta (\text{size} \times \text{invocation frequency} \times \text{memory\ energy\ unit)}}{\Delta \text{size}} \\
&= \ \text{invocation frequency} \times \text{memory\ energy\ unit} 
\end{aligned}
$}
\end{equation}
with the memory energy unit representing the energy required to load a byte from memory, we adopt 1.5pJ/byte for local memory and 120pJ/byte~\cite{On_Chip_Mem} for off-chip memory. Component size determines whether weights fit in energy-efficient local memory or power-hungry off-chip memory, influencing power sensitivity.

Accuracy sensitivity is calculated by progressively reducing a component's size, observing the effect on model accuracy, and fitting an exponential function to describe the relationship. The derivative of this function quantifies accuracy sensitivity.

Finally, we use the power-to-accuracy sensitivity ratio to prioritize compression decisions:
\begin{equation}
\scalebox{0.85}{$
\begin{aligned}
    \text{power-to-accuracy sensitivity ratio}  
     = \frac{\text{power sensitivity}}{\text{accuracy sensitivity}}
\end{aligned}
$}
\end{equation}

A higher ratio identifies components where compression provides the greatest power savings for minimal accuracy loss, helping determine the optimal compression order for on-device ASR models.

Our compression algorithm starts with a fully uncompressed model and iteratively reduces its size to achieve a user-defined power reduction target (e.g., "reduce power by 60 mW"). At each step, we calculate the power-to-accuracy sensitivity ratio for each component and compress the one with the highest ratio. In Neural Transducer models, the Joiner typically starts with the highest ratio due to its high power sensitivity from frequent invocation. Once its size is reduced enough to fit into energy-efficient local memory, its ratio decreases, and the Predictor becomes the next priority. The Predictor is compressed until it reaches its user-defined minimum size, beyond which further compression would cause significant accuracy loss due to the exponential relationship between accuracy and size. The Encoder is then compressed similarly, followed by additional compression of the Joiner if more power reduction is required.

 The compression order is thus: Joiner → Predictor → Encoder → Joiner. Our algorithm determines only the compression order between components, delegating the pruning of weight parameters within a selected component to existing compression methods. This makes our approach compatible with any existing compression algorithm.

\section{Experiments\label{sec:results}}

\subsection{Datasets and Models\label{subsec:data}}
We conduct experiments on two datasets: LibriSpeech and Public Video (details in Appendix~\ref{sec:appendix_dataset}).

LibriSpeech, from audiobooks, contains 960 hours of training data and two evaluation sets: Test-Clean, with easily transcribed recordings, and Test-Other, featuring recordings with accents or poor audio quality. Public Video, an in-house dataset of de-identified audio from publicly available English videos (with consent), includes 148.9K hours of training data and two evaluation sets: Dictation (5.8K hours of open-domain conversations) and Messaging (13.4K hours of audio messages).

For LibriSpeech, we use Emformer models~\cite{Emformer} with a 40ms input stride and 160ms chunk size. For Public Video, we use Conformer models~\cite{conformer} with a 60ms input stride and 300ms chunk size.

\subsection{Baselines and Evaluation Methodologies}
Our method identifies the most critical model component for compression to maximize energy savings. The specific compression technique applied to the identified component is beyond our scope.

We compare two scenarios: a uniform application of a baseline compression technique across the entire model ("baseline") and an enhanced version where the same technique is guided by our approach to strategically prioritize components ("baseline + our approach"). This comparison demonstrates the power savings achieved by our method and highlights the benefits of strategic component prioritization.

Our experiments use Adam-prune~\cite{yang2022omni}, the state-of-the-art compression technique for speech recognition models. While we employ the strongest available baseline, the choice or number of baselines is not critical, because our primary focus is on demonstrating consistent power savings achieved by integrating our approach with the baseline, irrespective of the baseline's inherent performance. Stronger baselines yield higher accuracy, and weaker baselines result in lower accuracy; however, the relative power savings for a given model size remain consistent. Therefore, the baseline selection does not affect our objective of highlighting power efficiency enhancement.



\begin{figure}
  \centering
  \begin{subfigure}[b]{0.48\linewidth}  
    \includegraphics[width=\linewidth]{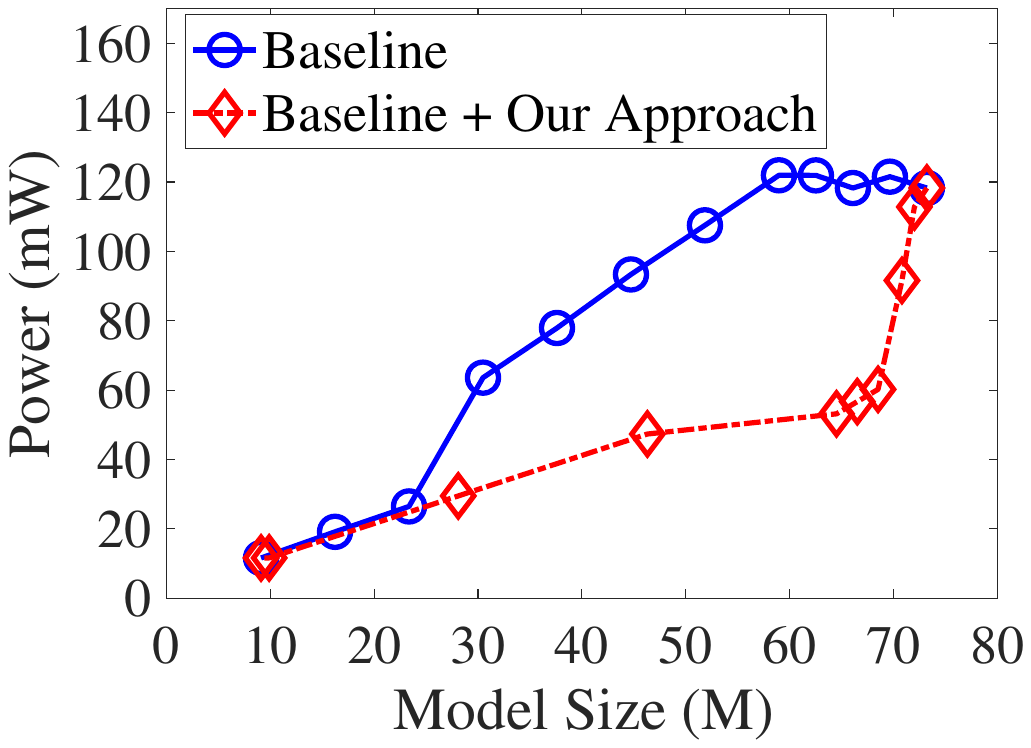}
    \caption{Power Consumption}
  \end{subfigure}
  \hfill 
  \begin{subfigure}[b]{0.48\linewidth} 
    \includegraphics[width=\linewidth]{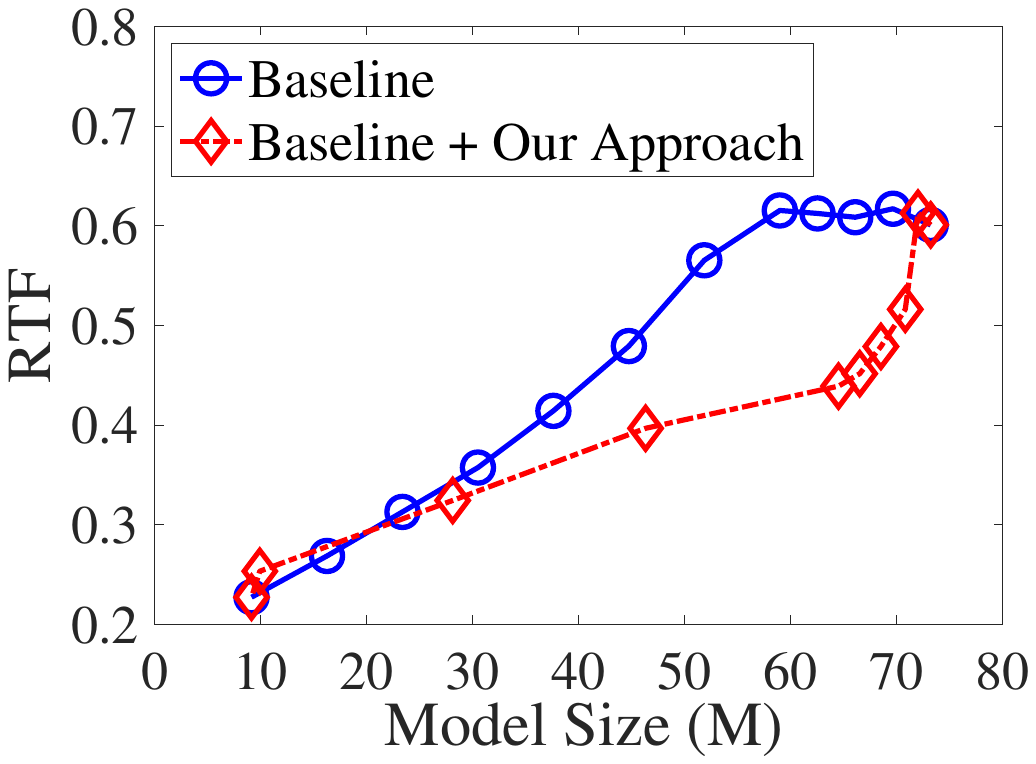}
    \caption{RTF}
  \end{subfigure}
  \newline 
  \begin{subfigure}[b]{0.48\linewidth} 
    \includegraphics[width=\linewidth]{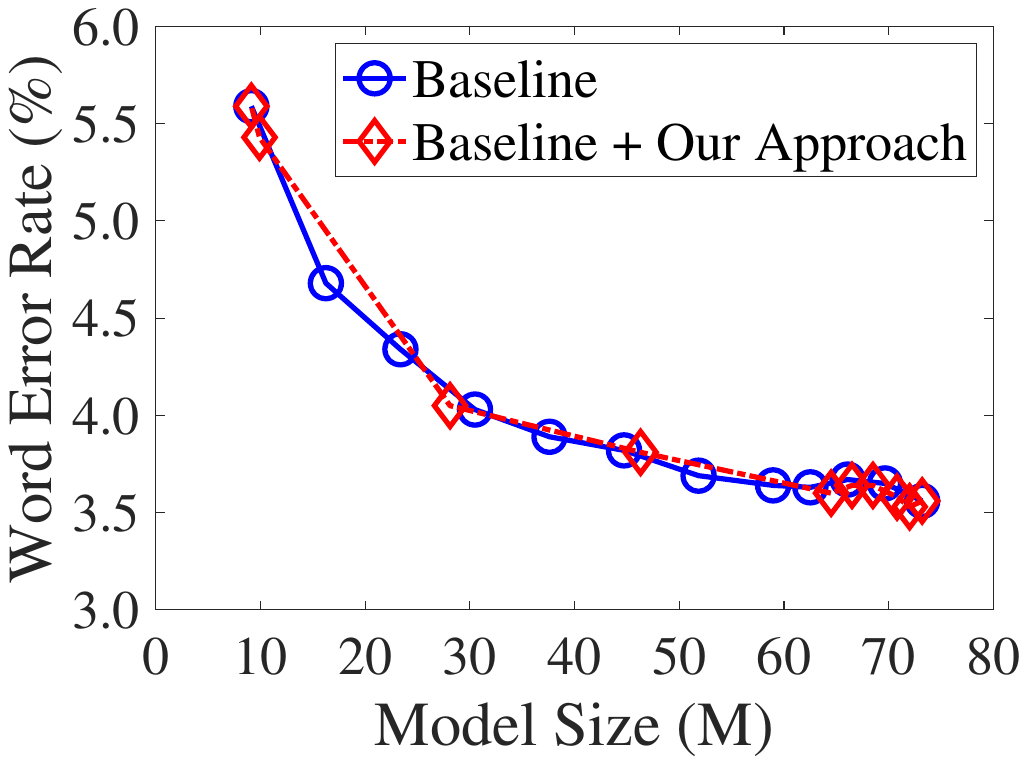}
    \caption{Error Rate on Test-Clean}
  \end{subfigure}
  \hfill 
  \begin{subfigure}[b]{0.48\linewidth} 
    \includegraphics[width=\linewidth]{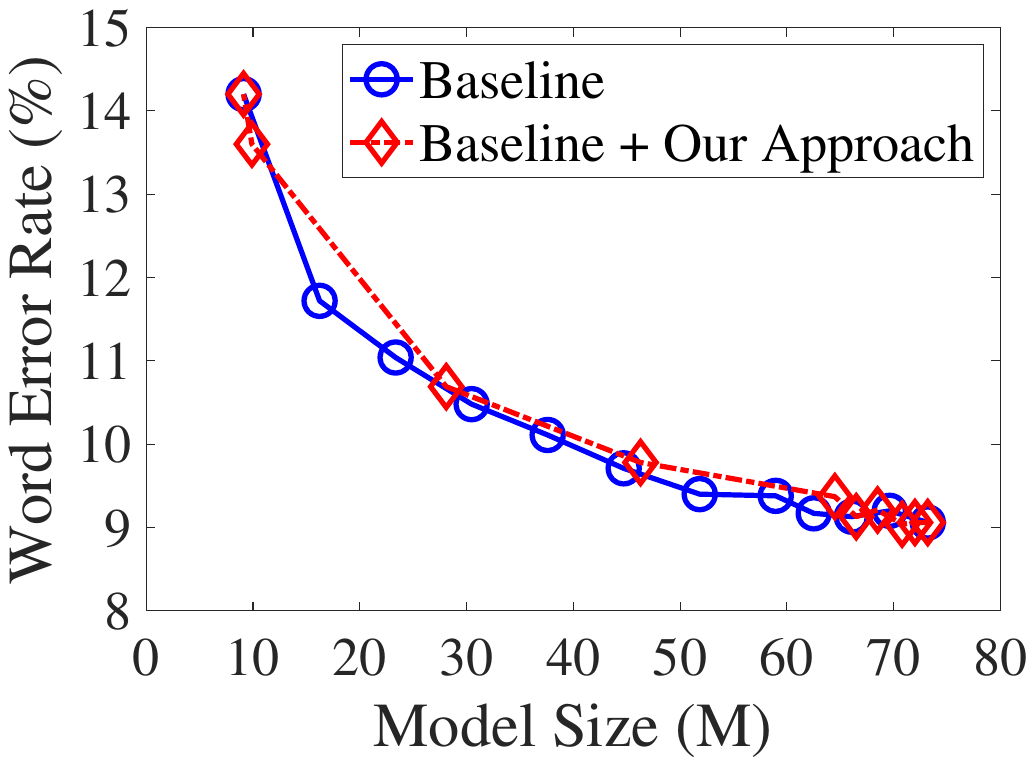}
    \caption{Error Rate on Test-Other}
  \end{subfigure}
  \caption{Models trained on LibriSpeech under different sizes and compression schemes.}
  \label{fig:libri_model}
\end{figure}
\subsection{Results on LibriSpeech \label{subsec:libri_results}}

Figure~\ref{fig:libri_model} (a) shows the power consumption across different model sizes. Our method achieves significant power savings compared to the baseline for models between 30–76 MB. For models under 30~MB, further compression results in minimal-size components, reducing differences between methods and leading to similar power consumption.

Figure~\ref{fig:libri_model} (b) illustrates the Real-Time Factor (RTF). Interestingly, while focusing on energy efficiency, our method improves RTF, indicating faster inference. This is due to prioritizing compression of heavily used components, which more significantly reduces overall inference time.

Figures~\ref{fig:libri_model} (c) and (d) show that word error rates remain consistent across model sizes, demonstrating that our method preserves baseline accuracy. Overall, Figures~\ref{fig:libri_model} (a)--(d) highlight that our approach reduces energy consumption by up to 47\% and RTF by 29\% while maintaining accuracy comparable to the baseline.

\subsection{Results on Public Video\label{subsec:product_results}}

\begin{figure}
  \centering
  \begin{subfigure}[t]{0.48\linewidth}  
    \includegraphics[width=\linewidth]{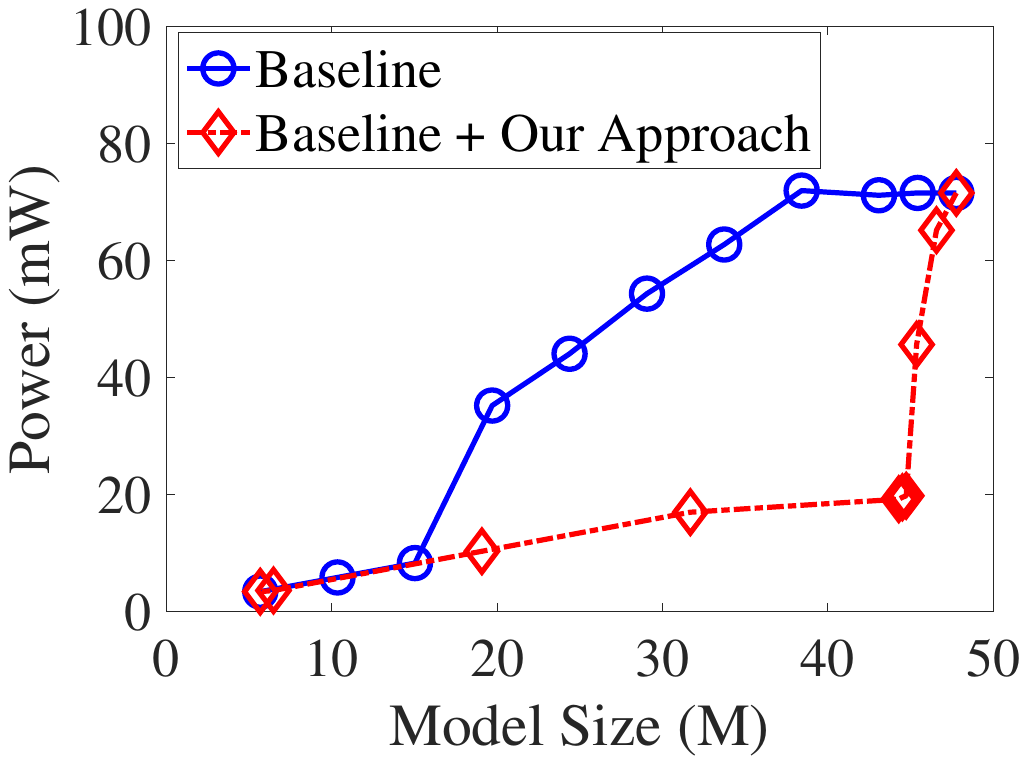}
    \caption{Power Consumption}
  \end{subfigure}
  \hfill 
  \begin{subfigure}[t]{0.48\linewidth} 
    \includegraphics[width=\linewidth]{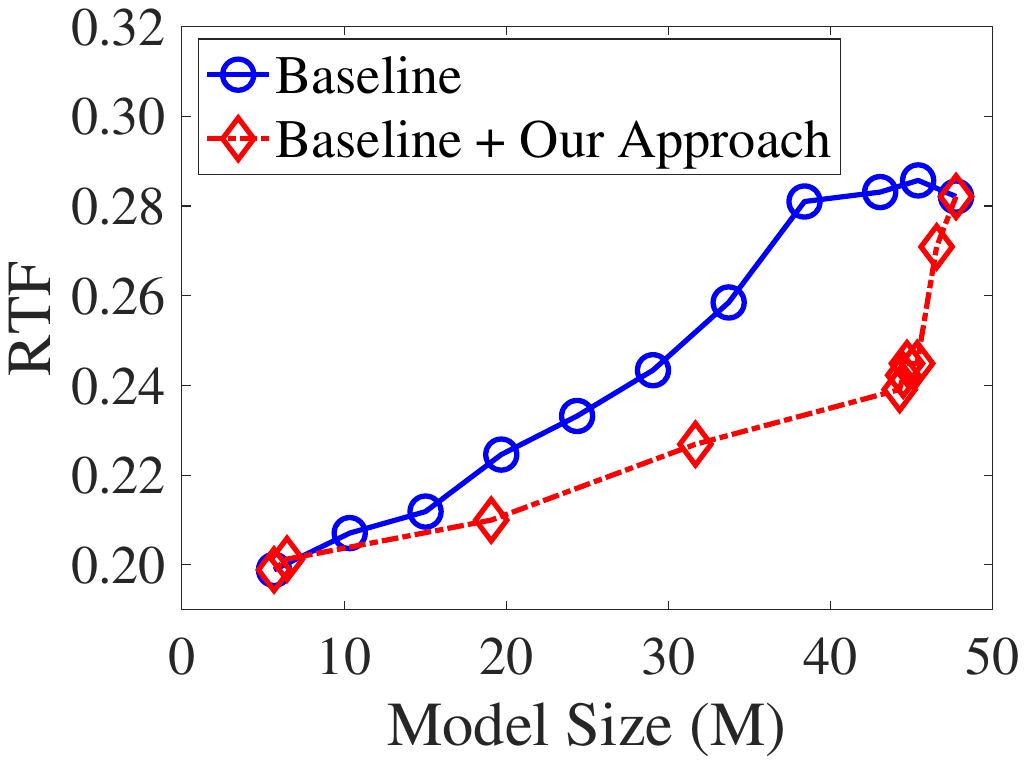}
    \caption{RTF}
  \end{subfigure}
  \newline 
  \begin{subfigure}[t]{0.48\linewidth} 
    \includegraphics[width=\linewidth]{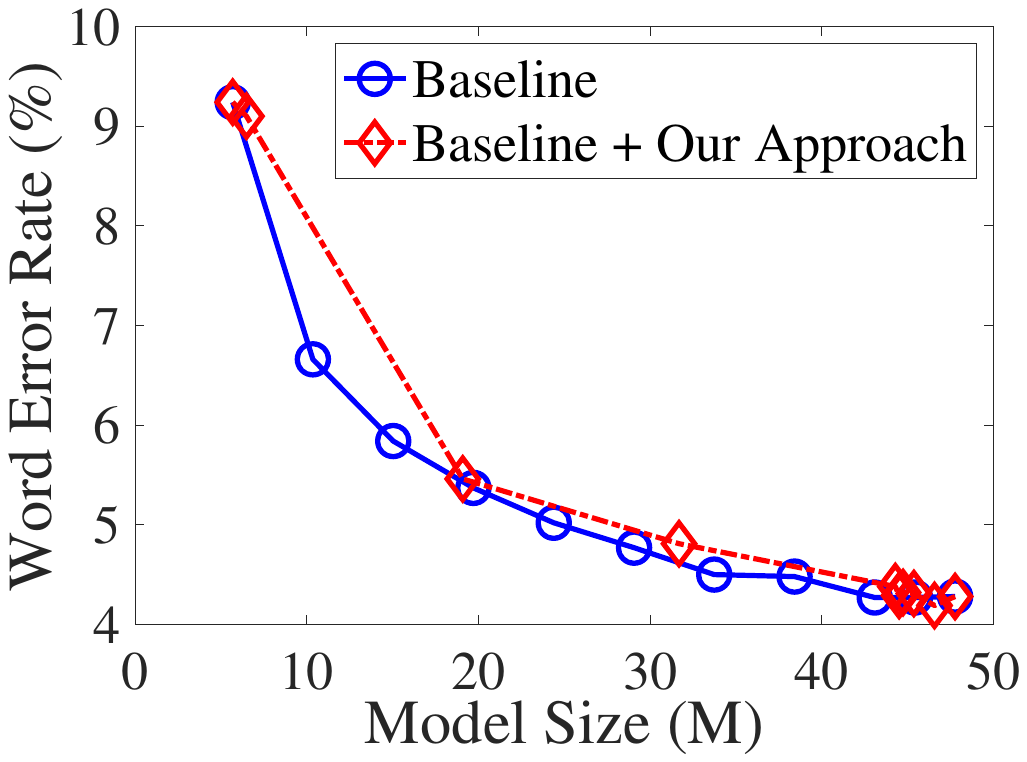}
    \caption{Error Rate on Messaging}
  \end{subfigure}
  \hfill 
  \begin{subfigure}[t]{0.48\linewidth} 
    \includegraphics[width=\linewidth]{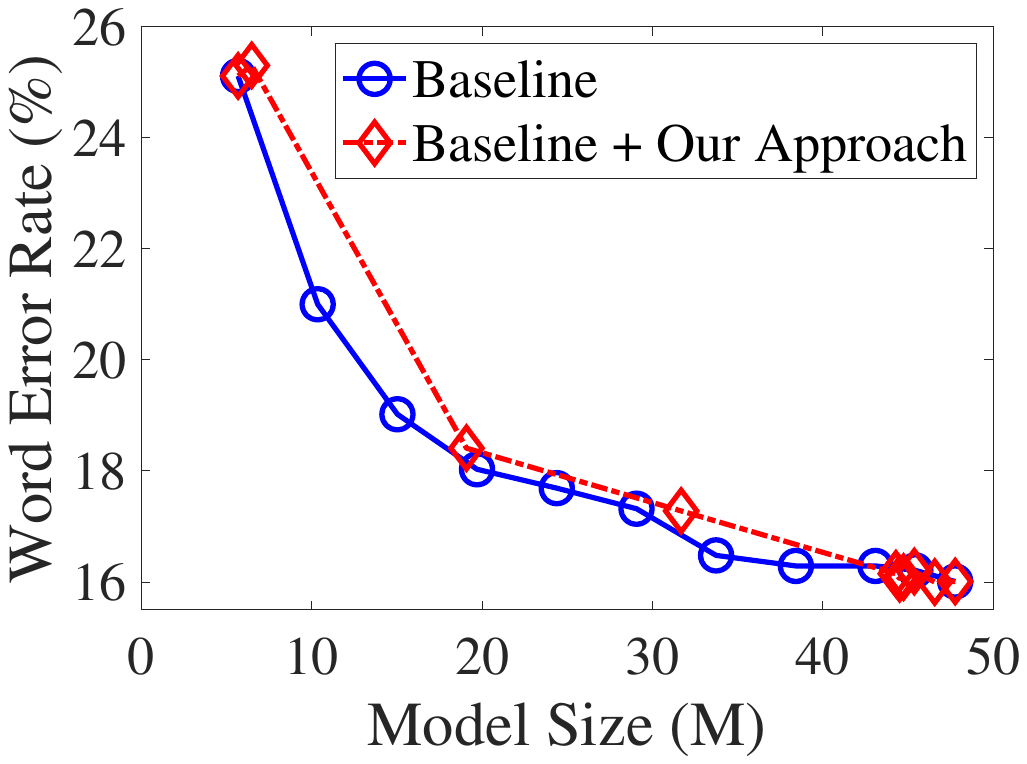}
    \caption{Error Rate on Dictation}
  \end{subfigure}
  \caption{Models trained on Public Video under different sizes and compression schemes.}
  \label{fig:product_model}
\end{figure}

Figures~\ref{fig:product_model} (a)--(d) show the power consumption, RTF, and accuracy for models of various sizes trained on the Public Video dataset. Our method reduces energy consumption by up to 38\% and RTF by 15\% while preserving accuracy.

\subsection{Discussion\label{subsec:discussion}}
As hardware technology advances, on-chip local memory in mobile and wearable devices continues to expand, allowing an increasing portion of Neural Transducer model weights to be stored locally. This shift enhances energy efficiency by leveraging the high energy efficiency of the on-chip memory. Simultaneously, these advancements may enable the deployment of more complex speech model architectures—previously infeasible for on-device or streaming scenarios due to model size and hardware constraints—as viable on-device streaming solutions. Consequently, we believe that power consumption will remain an important bottleneck in on-device streaming speech recognition. When new architectures incorporate multiple components with varying invocation frequencies, each component exhibits distinct power sensitivities. Our proposed energy efficiency optimization guidelines, which account for differences in power-to-accuracy sensitivity across model components, remain highly relevant in such cases. By adopting these guidelines, power consumption can be significantly reduced, fostering broader development, applicability, and deployment of on-device streaming speech recognition technology.

\section{Related Work\label{sec:related}}

This study is the first to analyze the operational dynamics and memory placement of model components to enhance energy efficiency in on-device streaming ASR. The most relevant prior works focus on ASR compression and power optimization.

\subsection{On-Device ASR Compression}
\citet{ghodsi2020rnn} demonstrated that removing recurrent layers from the Predictor in Neural Transducer models does not degrade word-error rates, enabling stateless operation and potential compression. \citet{botros2021tied} proposed parameter sharing between the Predictor and Joiner embedding matrices, introducing a weighted-average embedding to capture Predictor token history and reduce footprint. \citet{shangguan2019optimizing} reduced Predictor size by replacing LSTM units with sparsified Simple Recurrent Units (SRU) and adapted Encoders with sparsified CIFG LSTMs. \citet{yang2022omni} applied Supernet-based neural architecture search to optimize layer sparsity, balancing accuracy and size. While these works focused on reducing model size or RTF, they did not address power consumption, which is the central goal of our study.

\subsection{On-Device ASR Power Optimization}
Efforts to optimize Neural Transducer power consumption often involve modifying cell architectures. \citet{li2023folding} introduced folding attention, reducing model size and power consumption by 24\% and 23\%, respectively, without sacrificing accuracy. \citet{memoryEfficienRNNT} streamlined LSTM cells and designed a deeper, narrower model, reducing off-chip memory access by 4.5x and energy costs by 2x, with minimal accuracy loss. Our work differs by examining the runtime behaviors of Neural Transducer components to guide compression strategies specifically toward energy optimization.

\section{Conclusion\label{sec:conclusion}}\vspace{-0.5em}
Power consumption is a critical challenge for on-device streaming ASR, impacting device recharge frequency and user experience. This study analyzed power usage in ASR models, revealing its dependence on model size, invocation frequency, and memory placement. Notably, the Joiner consumes more power than the larger Encoder and Predictor due to its higher invocation frequency and off-chip memory usage. We also identified an exponential relationship between word error rate and encoder size.

Based on these insights, we developed guidelines for model compression to enhance energy efficiency. Applying these guidelines to the LibriSpeech and Public Video datasets achieved up to 47\% energy savings and a 29\% reduction in RTF, maintaining accuracy comparable to state-of-the-art methods. These findings highlight the potential of targeted optimizations to advance sustainable and energy-efficient on-device streaming speech recognition.



\bibliography{reference}
\balance
\appendix
\section{Details of Adam-Pruning Algorithm}
\label{sec:appendix_adamprune}
Adam-pruning is an iterative method designed to prune a model or its components. Each pruning step is executed over $N$ training epochs. During each step, Adam-pruning evaluates the square of the gradient ($E\left[\left(\frac{\partial l}{\partial w}\right)^2\right]$) for every non-sparse parameter $w$ in the model. A larger square of the gradient suggests that pruning the parameter would result in a substantial change in the model's performance. Based on this, Adam-pruning prunes only the parameters with the top $K$ smallest gradient squares at the end of each pruning step. After $M$ such steps, Adam-pruning reduces the model to a desired level of sparsity.

\section{Details of the Datasets}
\label{sec:appendix_dataset}
\subsection{LibriSpeech}
LibriSpeech~\cite{librispeech}, is a prominent corpus extensively utilized in speech recognition research. This corpus features 960 hours of English speech, sourced from audiobooks available through the LibriVox project, which are in the public domain. It includes two main evaluation sets tailored for different testing scenarios:
\begin{itemize}
    \item Test-Clean: This subset consists of high-quality, clean audio recordings. It provides an ideal condition for benchmarking the baseline performance of speech recognition systems due to its clarity and ease of transcription.
    \item Test-Others: This subset encompasses recordings that present a variety of challenges, such as accents, background noises, and lower recording qualities. It serves as a stringent testing environment to evaluate the robustness and adaptability of speech recognition technologies under less-than-ideal conditions.
\end{itemize}

\subsection{Public Video}
The Public Video dataset, an in-house collection, is derived from 29.8K hours of audio extracted from English public videos. 
This dataset has been ethically curated with the consent of video owners and further processed to ensure privacy and enhance quality. We de-identify the audio, aggregate it, remove personally identifiable information (PII), and add simulated reverberation. We further augment the audio with sampled additive background noise extracted from publicly available videos. Speed perturbations~\cite{ko2015audio} are applied to create two additional copies of the training dataset at 0.9 and 1.1 times the original speed. We apply distortion and additive noise to the speed-perturbed data. These processing steps eventually result in a total of 148.9K hours of training data. For evaluating the performance of models trained on this dataset, we use the following two test sets:
\begin{itemize}
    \item Dictation: This subset consists of 5.8K hours of human-transcribed, anonymized utterances, sourced from a vendor. Participants were asked to engage in unscripted open-domain dictation conversations, recorded across various signal-to-noise ratios (SNR), providing a diverse assessment environment.
    \item Messaging: This subset comprises 13.4K hours of utterances, sourced from a vendor. It features audio messages recorded by individuals following scripted scenarios intended for an unspecified recipient. These utterances are generally shorter and incorporate more noise than those in the dictation subset, offering a different dimension to evaluate ASR systems.
\end{itemize}

\section{Accuracy of ASR Models Trained on Public Video}
\label{sec:appendix_product_asr}
\begin{figure}[h]
    \begin{subfigure}{.45\linewidth}
    \includegraphics[width=\linewidth]{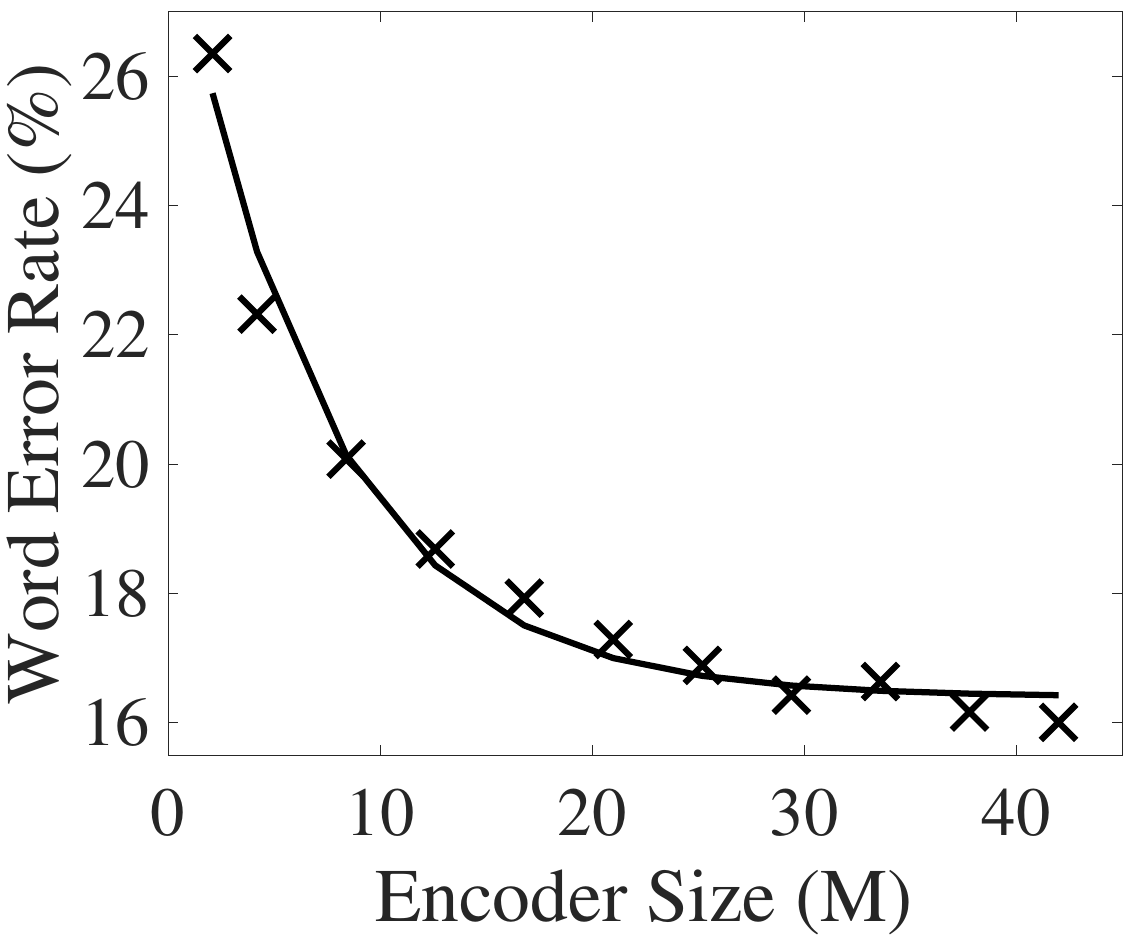}
    \caption{Dictation}
    \end{subfigure}
    \hspace{0.5cm}
    \begin{subfigure}{.45\linewidth}
    \includegraphics[width=\linewidth]{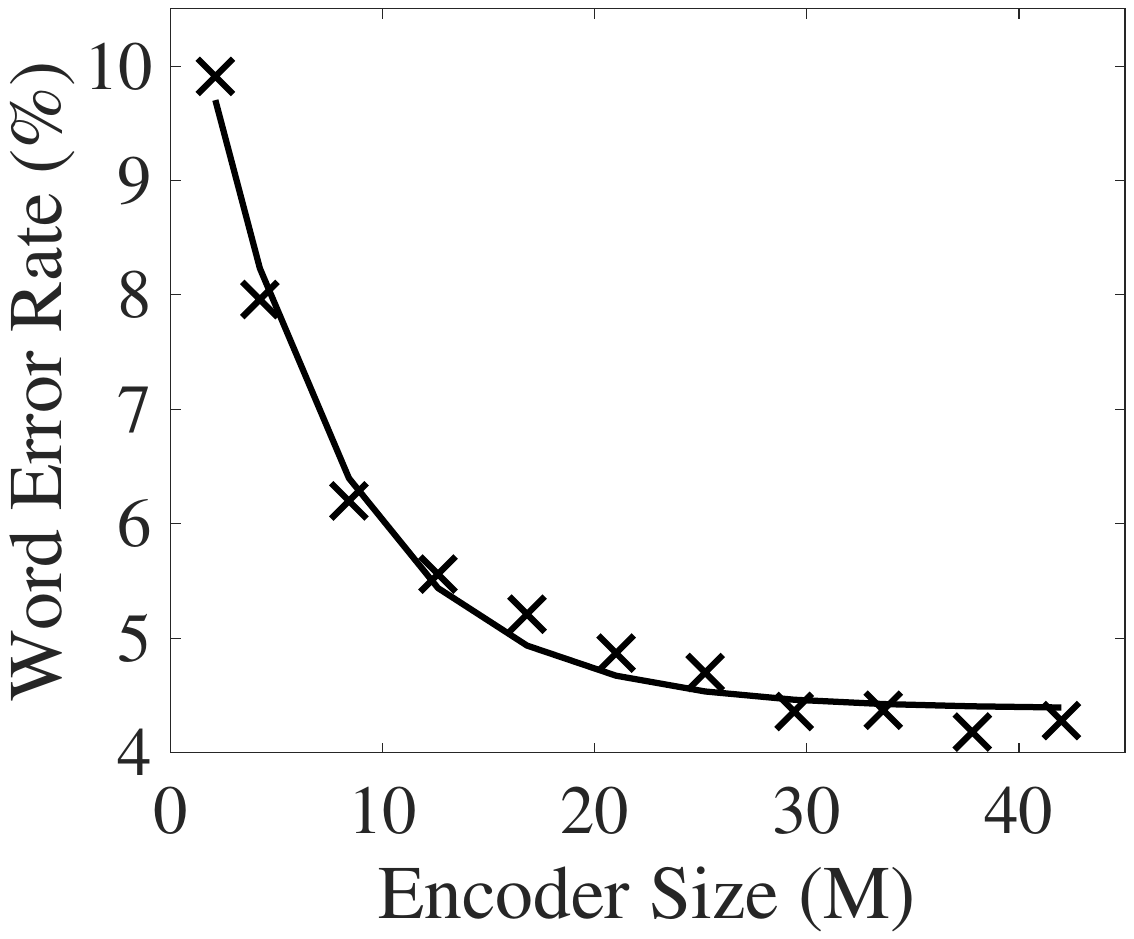}
    \caption{Messaging}
    \end{subfigure}
    \caption{Models trained on the Public Video dataset: Word error rate with compressing Encoder while keeping the size of Predictor and Joiner.}
    \label{fig:product_encoder_dictation}
\end{figure}

We applied compression to the Encoder of the ASR model trained using the Public Video dataset. The impact of this compression on word error rates across two evaluation sets, Dictation and Messaging, is depicted in Figures~\ref{fig:product_encoder_dictation} (a) and (b). To analyze the data, we employed the function outlined in Equation~\ref{eq:1}, which proved to be an excellent fit; the predictions derived from this function align closely with the observed data. Quantitatively speaking, the adjusted R-squared values—0.9760 for Dictation and 0.9851 for Messaging—underscore the exponential relationship between word error rate and encoder size, reaffirming this pattern's consistency across different datasets.

\section{Impact of Input Stride and Chunk Size on Model Accuracy and Power Usage}
\label{sec:appendix_param_sweep}
Input stride and chunk size are two essential hyperparameters for on-device streaming ASR. Input stride defines the time window over which input frames are combined into an aggregated frame that is then fed into the model. Chunk size refers to the time duration over which these aggregated frames are processed together as a batch by the model. In this section, we examine how varying these parameters affects the performance and power consumption of the Neural Transducer.

We first vary the input stride from 20 milliseconds to 40 milliseconds and evaluate the accuracy and power consumption of four models trained on LibriSpeech: a dense model, a model with 80\% sparsity in its encoder, a model with 80\% sparsity in its predictor, and a model with 80\% sparsity in its joiner. The results are provided in Tables~\ref{tab:appendix stride accuarcy} and~\ref{tab:appendix stride power}.

\begin{table*}[t]
\centering 
\resizebox{0.9\textwidth}{!}{
\renewcommand{\arraystretch}{1.4}
\begin{tabular}{cccccc}
\hline
 \textbf{\makecell{Word Error Rate \\(\%)}} & Input Stride & Dense Model & \makecell{80\% Sparse\\ Encoder} & \makecell{80\% Sparse\\ Predictor} & \makecell{80\% Sparse\\ Joiner} \\ \hline
Test-Clean & 20ms & 3.61 & 4.72 & 3.61 & 4.17 \\ 
                                     & 40ms & 3.56 & 4.86 & 3.60 & 3.64 \\ \hline
Test-Other & 20ms & 9.13 & 11.90 & 9.13 & 9.58 \\ 
                                     & 40ms & 9.06 & 12.08 & 9.14 & 9.29 \\ \hline
\end{tabular}}
\caption{Impact of input stride on the model accuracy trained on LibriSpeech.\label{tab:appendix stride accuarcy}}
\end{table*}

\begin{table*}[t]
\centering
\resizebox{0.9\textwidth}{!}{
\renewcommand{\arraystretch}{1.4}
\begin{tabular}{cccccc}
\hline
 \textbf{\makecell{Model Power\\ Consumption (mW)}} & Input Stride & Dense Model & \makecell{80\% Sparse\\ Encoder} & \makecell{80\% Sparse\\ Predictor} & \makecell{80\% Sparse\\ Joiner} \\ \hline
 & 20ms & 131 & 104 & 123 & 62 \\
 & 40ms & 118 & 92 & 110 & 62  \\ \hline
\end{tabular}}
\caption{Impact of input stride on the power consumption of models trained on LibriSpeech.\label{tab:appendix stride power}}
\end{table*}

\begin{table*}[t]
\centering
\resizebox{0.9\textwidth}{!}{
\renewcommand{\arraystretch}{1.4}
\begin{tabular}{cccccc}
\hline
 \textbf{\makecell{Word Error Rate \\(\%)}} & Chunk Size & Dense Model & \makecell{80\% Sparse\\ Encoder} & \makecell{80\% Sparse\\ Predictor} & \makecell{80\% Sparse\\ Joiner} \\ \hline
Test-Clean & 160ms & 3.56 & 4.86 & 3.60 & 3.64 \\ 
                                     & 320ms & 3.50 & 4.60 & 3.50 & 3.52 \\ \hline
Test-Other & 160ms & 9.06 & 12.08 & 9.14 & 9.29 \\ 
                                     & 320ms & 8.82 & 11.75 & 8.83 & 8.90 \\ \hline
\end{tabular}}
\caption{Impact of chunk size on the model accuracy trained on LibriSpeech.\label{tab:appendix chunk accuarcy}}
\end{table*}

\begin{table*}[t]
\centering
\resizebox{0.9\textwidth}{!}{
\renewcommand{\arraystretch}{1.4}
\begin{tabular}{cccccc}
\hline
 \textbf{\makecell{Model Power\\ Consumption (mW)}} & Chunk Size & Dense Model & \makecell{80\% Sparse\\ Encoder} & \makecell{80\% Sparse\\ Predictor} & \makecell{80\% Sparse\\ Joiner} \\ \hline
 & 160ms & 118 & 92 & 110 & 62 \\
 & 320ms & 94 & 86 & 87 & 38  \\ \hline
\end{tabular}}
\caption{Impact of chunk size on the power consumption of models trained on LibriSpeech.\label{tab:appendix chunk power}}
\end{table*}

\FloatBarrier

\noindent Our findings are as follows:
\begin{itemize}
    \item Observation 1: A smaller stride can have both positive and negative effects on model performance.
    \item Observation 2: A smaller stride generally increases power consumption.
\end{itemize}

Regarding the first observation, input stride is used to enhance training and inference efficiency by reducing sequence length. While a smaller stride better preserves local acoustic features and improves performance, it also introduces risks such as greater sensitivity to noise and loss of broader contextual information. A stride of 4--6 is commonly chosen to balance accuracy and efficiency.

As for the second observation, in streaming ASR, a smaller stride increases the number of segments, resulting in more frequent decoding of blank tokens and thus more frequent invocation of the joiner, which raises power consumption. However, if the joiner is compressed to fit within the SRAM, this increased invocation has minimal impact on power usage, due to the high energy efficiency of SRAM.

We also vary the chunk size from 160ms to 320ms and measure the accuracy and power consumption of four models: a dense model, a model with 80\% sparsity in its encoder, a model with 80\% sparsity in its predictor, and a model with 80\% sparsity in its joiner. The results are provided in Tables \ref{tab:appendix chunk accuarcy} and~\ref{tab:appendix chunk power}. Our observations are as follows:
\begin{itemize}
\item Observation 3: Increasing the chunk size generally improves model accuracy.
\item Observation 4: Larger chunk sizes reduce model power consumption.
\end{itemize}
For the third observation, larger chunk sizes enable the encoder to capture relationships between segments more effectively, improving performance. However, smaller chunk sizes have the advantage of lowering decoding latency.

As for the fourth observation, in streaming ASR, a larger chunk size decreases the frequency at which the encoder is invoked, thereby reducing memory power usage and overall power usage.

\end{document}